# Propositional computability logic I

Giorgi Japaridze[*]

Department of Computing Sciences, Villanova University, 800 Lancaster Avenue, Villanova, PA 19085, USA.
Email: giorgi.japaridze@villanova.edu URL: http://www.csc.villanova.edu/~japaridz/

**Abstract**

In the same sense as classical logic is a formal theory of truth, the recently initiated approach called *computability logic* is a formal theory of computability. It understands (interactive) computational problems as games played by a machine against the environment, their computability as existence of a machine that always wins the game, logical operators as operations on computational problems, and validity of a logical formula as being a scheme of "always computable" problems. Computability logic has been introduced semantically, and now among its main technical goals is to axiomatize the set of valid formulas or various natural fragments of that set. The present contribution signifies a first step towards this goal. It gives a detailed exposition of a soundness and completeness proof for the rather new type of a deductive propositional system **CL1**, the logical vocabulary of which contains operators for the so called *parallel* and *choice* operations, and the atoms of which represent *elementary problems*, i.e. predicates in the standard sense.

This article is self-contained as it explains all relevant concepts. While not technically necessary, however, familiarity with the foundational paper "Introduction to computability logic" [Annals of Pure and Applied Logic 123 (2003), pp.1-99] would greatly help the reader in understanding the philosophy, underlying motivations, potential and utility of computability logic — the context that determines the value of the present results.

## 1 Introduction

The approach initiated recently in [7] called *computability logic* is a formal theory of computability in the same sense as classical logic is a formal theory of truth. It understands (interactive) *computational problems* as games played by a machine against the environment, their computability as existence of a machine that always wins the game, logical operators as operations on computational problems, and validity of a logical formula as being a scheme of "always computable" problems. With the meaning of *computational resources* being dual/symmetric to that of computational problems (what is a problem for the machine is a resource for the environment, and vice versa), computability logic is, at the same time, a logic of computational resources.

While the original motivations behind this approach were theoretical, more than one expert has noted that computability logic essentially "competes for the same market as linear logic". Such an assessment apparently does have a point, and simplifies the job of explaining the significance of the approach: most of the reasons that attract the attention of computer scientists to linear logic apply to computability logic as well. The general philosophies of linear and computability logics are indeed similar in their striving to develop a logic of resources. The ways this philosophy is materialized, however, are rather different. The scheme that linear logic has been following can be characterized as '*from syntax to semantics*', whereas the motto of computability logic is '*from semantics to syntax*' (by syntax here we mean an axiomatic/deductive

[*]This material is based upon work supported by the National Science Foundation under Grant No. 0208816

construction of a logic). The natural way to develop a new logic is to first (1) present a philosophy, motivation and intuitions, i.e. an informal semantics; then (2) elaborate a formal semantics that adequately captures those intuitions, and only after that (3) ask what the corresponding logic (syntax) is. The way classical logic evolved and climaxed in Gödel's completeness theorem matches this pattern. And this is exactly the scheme that computability logic follows. On the other hand, looking back at the history of linear or intuitionistic logics, one can see that they essentially jumped from step (1) directly to step (3). That is, a particular deductive system was proposed (step (3)) based on some general intuitions (step (1)) without having a mathematically strict and intuitively convincing formal semantics,[1] so that, in the absence of a clear concept of truth or validity, the question regarding whether the proposed system was complete and hence "right", could not even be meaningfully asked. Only retroactively have attempts been made to find the missing formal semantics. Technically it is always possible to come up with some sort of a semantics that matches a given target syntactic construction, but the whole question is how natural and meaningful such a semantics is and how adequately it captures the original intuitions and motivations (step (1)) underlying the logic. Unless the target construction, by good luck, really *is* "the right logic", the chances of a decisive success when following the odd scheme 'from syntax to semantics' could be rather slim.

Mentioning intuitionistic logic here is no accident. What has been said about computability vs. linear logic also extends to computability vs. intuitionistic logic. The former offers a meaningful semantics that has good claims to adequately capture the main intuitions associated with the latter, in particular, its constructivistic intuitions and Kolmogorov's known thesis according to which intuitionistic logic is a logic of problems. One of the main conjectures stated in [7] is that Heyting's intuitionistic calculus is sound and complete with respect to computability-logic semantics. On the other hand, linear logic or even affine logic (linear logic + weakening) can be shown to be sound but incomplete.

Classical logic is another natural fragment of computability logic. This is due to the fact that, as we are going to see, classical predicates are nothing but a special sort of computational problems — problems with zero degree of interactivity — and the classical concept of truth is nothing but our concept of computability restricted to this special sort of problems. Thus, computability logic, rather than defying classical logic, simply refines, generalizes and conservatively extends it, which makes the former a reasonable and appealing alternative to the latter in virtually every aspect of its applications. In particular, applied theories — such as, say, Peano arithmetic — can be based on computability logic instead of classical logic, getting this way much more expressive, constructive and computationally meaningful theories than their classical-logic-based counterparts. In a similar sense, computability logic can be considered a refinement, generalization and extension of intuitionistic and linear (the latter understood in a very broad sense) logics, which makes it a universal framework that integrates — on the basis of one semantics — the above three popular logics/traditions with their seemingly unrelated or even antagonistic philosophies.

Taking into account that most real computing systems are interactive and that computability logic as a game logic is in fact a logic of interaction, the latter has potential applications in various areas of computer science, including theory of interactive computation, protocols, knowledgebase systems, or resource-based planning systems. [7, 8] contain ample discussions and illustrations.

Among the main technical goals of computability logic — just as of any other semantically introduced logic — is to axiomatize the set of valid formulas or various natural fragments of it. The present contribution signifies a first step towards this goal. It gives a detailed exposition of a soundness and completeness proof for the rather new/unusual type of a deductive propositional system **CL1**, the logical vocabulary of which contains operators for the so called *parallel* and *choice* operations, and the atoms of which represent the above-mentioned "special sort" of problems, i.e. predicates.

It is not an ambition of the present paper to motivationally (re)introduce and (re)justify computability logic. This job has been done in [7] and, in a more compact and less technical way, in [8]. The author assumes that most likely the reader is familiar with at least the motivational/philosophical parts of either article and that this is why he/she has decided to read the present paper. Yet, this paper is mathematically self-contained in that it defines all relevant concepts.

---

[1] A reader who shares the author's taste would easily understand why the phase, coherent, Kripke or other technical semantics do not count here.

# 2 Main concepts explained informally

Computability is a property of *computational problems*, and the first basic issue to be clarified here is what "computational problem" means. Apparently any answer to this question will match the general scheme according to which a computational problem is a binary relation $R$ between *possible inputs* and *possible outputs*, with $R(e,\Gamma)$ meaning that $\Gamma$ is a "right output" on input $e$. That is, a computational problem is the task of generating, for any possible input $e$, an output $\Gamma$ that satisfies $R(e,\Gamma)$. Here input is understood as something that is given from the very beginning and remains fixed throughout the computation, and obviously there is no loss of generality in assuming that it is always a (possibly infinite) string over some finite alphabet — or, equivalently, is a tuple or a sequence of natural numbers. The question about the type of the output, and how exactly it is generated, is trickier. In the traditional Church-Turing approach — which, incidentally, often insists that input be finite and $R$ be functional — output is just a (finite) string, and the computing agent has full control over its generation, with the outside world being shut out throughout the computation. Such an understanding, however, is too narrow as this has been already acknowledged by the theoretical computer science community [4, 10, 12] and apparently by Turing [11] himself. Most of the tasks that real computers perform are interactive, where both the computing system and its environment may remain active throughout the computation and communicate with each other through what is called "*observable actions*" [10, 12]; then, what constitutes the output is not just a simple value generated by the computing system at the end of the process (which can be thought of as one single observable action by the system), but rather the entire *interaction history*, — the potentially long or even infinite sequence of the observable actions taken by the interacting parties, so that the system now only has partial control over the observable outcome (output) of the process. The task performed by a server, which essentially is an infinite dialogue with the environment (clients),[2] perhaps is a good example of a computational problem not captured by the traditional, non-interactive models.

As this was mentioned, technically we understand interactive problems as *games*, or *dialogues*, between two agents/players: *machine* and *environment*, symbolically named as $\top$ and $\bot$, respectively. Machine, as its name suggests, is specified as a mechanical device with fully determined, algorithmic behavior, while the behavior of the environment, that represents a capricious user or the blind forces of nature, is allowed to be arbitrary. "Observable actions" by these two agents translate into game-theoretic terms as their *moves*, "interaction histories" as *runs*, i.e. sequences of moves, "possible outputs" as *legal runs*, and "right outputs" as *successful*, or *winning runs*. The machine is considered to solve a given problem/game if it wins it (ensures that a right/successful output/run is generated) on every possible input no matter how the environment behaves.

*Predicates* in their classical sense are thought of as a special sort of computational problems called *elementary*. These are games without any moves, i.e. "dialogues" of length 0. The machine automatically wins or loses an elementary game $A$ on input $e$ depending on whether $A(e)$ is true or false (true = won, false = lost). Another special sort of computational problems is what we call *constant games*. These are games that do not depend on input. Obviously then classical propositions (0-ary predicates) are nothing but constant elementary games, so that there are exactly two constant elementary games: $\top$, won by $\top$, and $\bot$, won by $\bot$.

The game of chess is a good intuitive example of a constant non-elementary game, and helpful in visualizing our basic game operations. Let us call this game — as it is seen by the white player — *Chess*. We assume that there are no draw outcomes in *Chess*: say, if the white player does not win within 100 moves, he is considered the loser.

The operation $\neg$, called *negation*, switches the roles in the game: $\top$'s moves and wins become $\bot$'s moves and wins, and vice versa. Thus, $\neg$*Chess* is nothing but the game of chess from the point of view of the black player. Notice that as elementary games have no moves, role switching there simply means switching the (automatic) winners. In other words, $\neg$ restricted to elementary games acts exactly as classical negation: it turns won (true) elementary games (predicates) into lost (false) elementary games (predicates), and vice versa.

The operations $\wedge$ and $\vee$ combine games in the following intuitive and technical sense: playing $A \wedge B$ or $A \vee B$ means playing the two games $A$ and $B$ in parallel. In $A \wedge B$ the machine is considered the winner if it wins in both of the components, while in $A\vee B$ winning in just one component is enough. This informal explanation might already be sufficient to guess that $\neg A\vee A$ is an always-winnable-by-machine combination of games, i.e. a valid principle of computability. Say, a simple winning strategy for $\neg Chess \vee Chess$ is to mimic, in *Chess*, the moves made by the environment in $\neg Chess$, and vice versa. However, not all classical principles will be retained. E.g., one can show that $\neg A\vee (A\wedge A)$ is not valid. While proving this might require some thought, at least it is clear that the above 'mimicking strategy' is no longer applicable: the best that the machine can do in the three-board game $\neg Chess \vee (Chess \wedge Chess)$ is to pair $\neg Chess$ with one of the two conjuncts of $Chess\wedge Chess$. But in that case it may happen that $\neg Chess$ and the unmatched *Chess* are both lost, which would make the whole game also lost.

---

[2]With, say, startup-time settings and allocated resources such as databases acting in the role of input.

This reminds us of linear logic which, too, rejects the principle $\neg A\vee (A\wedge A)$ with $\vee,\wedge$ understood as multiplicatives. While the behavior of $\neg,\wedge,\vee$ is indeed "similar" to that of the corresponding linear-logic operators, it is definitely not the same as it validates a properly bigger class of formulas. One of the principles that separate computability logic from linear or even affine logic is Blass's [3] principle

$$\bigl((\neg P\vee \neg Q)\wedge (\neg R\vee \neg S)\bigr) \vee \bigl((P\vee R)\wedge (Q\vee S)\bigr).$$

We call $\wedge$ and $\vee$ *parallel operations*. Technically the unary operation of negation $\neg$ also counts as a parallel operation in this paper. Just as this was the case with $\neg$, when applied to elementary games, $\wedge$ and $\vee$ again generate elementary games and, with the correspondence 'true=won, false=lost' in mind, mean virtually the same as classical conjunction and disjunction. This full conservation of classical meaning, that also extends to the quantifier level (see [7, 8]), is exactly what makes classical logic a natural fragment of computability logic: the former is nothing but the latter restricted to elementary problems.

The next important group of operations is what we call *choice operations*: $\sqcap$ and $\sqcup$, resembling (but again, not "the same as") the additive conjunction and disjunction of linear logic. $A_1\sqcup A_2$ is the game where, at the beginning, the machine has to make one of the moves '1' or '2', signifying a choice of component. After such a move $i$ is made, the game continues — and the winner is determined — according to the rules of the chosen disjunct $A_i$. In case the machine fails to make an initial move/choice, it is considered to have lost the game. $A_1\sqcap A_2$ is defined in the same way, with the only difference that here it is $\bot$ rather than $\top$ who makes an initial choice and who loses if such a choice is never made. The logical behavior of parallel operations is dramatically different from that of their choice counterparts. While $A\vee \neg A$ is valid, $A\sqcup \neg A$ is not, reminding us of both linear and intuitionistic logics. Back to our chess example, unlike $Chess\vee \neg Chess$, winning $Chess\sqcup \neg Chess$ is not easy: here, at the very beginning, the agent has to choose between *Chess* and $\neg Chess$, and then win the chosen one-board game.

The problem of deciding a predicate $P$ can now be expressed as the game $P\sqcup \neg P$. Solving/winning this problem/game means to choose, for any given input $e$, the $\sqcup$-disjunct that is true at $e$. Obviously this can be done by a machine if and only if $P$ is decidable. Hence, the failure of the principle $P\sqcup \neg P$ means nothing but existence of undecidable problems. Speaking philosophically, $\sqcup$ and $\sqcap$ are *constructive* versions of OR and AND.

*Parallel implication* $\rightarrow$ is among the most meaningful operations. Formally $A\rightarrow B$ is defined as $(\neg A)\vee B$. Intuitively $\rightarrow$ is a problem reduction operation: solving $A\rightarrow B$ means solving $B$ having $A$ as an (external) *computational resource*; in other words, it means *reducing $B$ to $A$*. To get a feel of this, let us borrow an example from [7] about reducing the acceptance problem to the halting problem. The former can be expressed as $A\sqcup \neg A$, with $A$ being the predicate "Turing machine $x$ accepts input $y$". And the latter can be expressed as $H\sqcup \neg H$, with $H$ meaning "Turing machine $x$ halts on input $y$". While the acceptance problem is not decidable, it is algorithmically reducible to the halting problem. In particular, there is a machine that always wins the game

$$(H\sqcup \neg H)\rightarrow (A\sqcup \neg A).$$

Here is a strategy for such a machine on input $(x,y)$: Wait till $\bot$ selects one of the $\sqcup$-disjuncts in the antecedent (where the roles of the two players are switched and hence $\sqcup$ signifies $\bot$'s choice; failure of $\bot$ to make such a choice results in $\top$'s winning the game). If $\bot$ selects $\neg H$ there, then select $\neg A$ in the consequent and celebrate victory. Otherwise, if $\bot$ chooses $H$, simulate machine $x$ on input $y$ until it halts. If the simulation accepts, select $A$ in the consequent, and if the simulation rejects, select $\neg A$ there. It is a possibility that this simulation goes on forever and hence no move will be made in the consequent; however, in such a scenario $H$ is false, and having "lied" in the antecedent makes $\bot$ the loser no matter what happens in the consequent.

The above strategy indeed is a reduction: it solves the consequent using an external solution to the antecedent. [7] argued that $\rightarrow$ adequately captures our strongest intuition of reducing one interactive problem to another. The weakest intuition of reduction is captured by another computability-logic operator $\circ\!\!-$ (denoted $\Rightarrow$ in [8]), whose logical behavior is conjectured to be exactly that of the intuitionistic implication. Restricted to non-interactive problems, $\circ\!\!-$ has been shown to be the same as Turing reduction — the sort of reduction that, unlike $\rightarrow$, allows uncontrolled usage ("recycling") of computational resources.

Several other basic operations have been semantically introduced within the framework of computability logic. They include: the quantifier-level counterparts of our conjunctions and disjunctions, such as $\sqcap$, with $\sqcap xA(x)$ meaning the infinite conjunction $A(0) \sqcap A(1) \sqcap A(2) \sqcap \ldots$; the so called *blind operations* that generate imperfect-information games; the propositional operations, called *recurrence operations*, that generate multiple copies of games and thus resemble the exponential operators of linear logic, etc. Their definitions can be found in [7, 8]. Our present paper is only focused on the $(\neg, \wedge, \vee, \rightarrow, \sqcap, \sqcup)$-fragment.

## 3 Interactive computational problems

To define our games formally, we need some technical terms and conventions. First of all, let us agree that by an **input**[3] we mean an infinite sequence of natural numbers. In most cases, of course, we would be interested in finite inputs. Every finite input can be understood as a tuple of natural numbers or — for known reasons — just as a single natural number. Our stipulation that input should be infinite, the only reason for which is generality and flexibility, creates no problem: of course every finite input $(a_1, \ldots, a_n)$ can be thought of as the infinite input $\langle a_1, \ldots, a_n, 0, 0, 0, \ldots \rangle$ and vice versa.

We generously define a **move** as any finite string over the standard keyboard alphabet. One of the non-numeric and non-punctuation symbols of the alphabet, denoted ♠, is designated as a special-status move, intuitively meaning a move that is always illegal to make. A **labeled move** is a move prefixed with $\top$ or $\bot$, with its prefix (**label**) indicating which player has made the move. A **run** is a (finite or infinite) sequence of labeled moves, and a **position** is a finite run.

We will be exclusively using letters $\Gamma, \Delta$ as metavariables for runs, $\Phi, \Psi, \Theta$ for positions, $\wp$ for players, $e$ for inputs, and $\alpha, \beta$ for moves. Runs will be often delimited with "$\langle$" and "$\rangle$", with $\langle\rangle$ thus denoting the **empty run**. The meaning of an expression such as $\langle \Phi, \wp\alpha, \Gamma \rangle$ must be clear: this is the result of appending to $\langle \Phi \rangle$ $\langle \wp\alpha \rangle$ and then $\langle \Gamma \rangle$.

**Definition 3.1** A **game** is a pair $A = (\mathbf{Lr}^A, \mathbf{Wn}^A)$, where:

- $\mathbf{Lr}^A$ is a function that sends every input $e$ to a set $\mathbf{Lr}^A_e$ of runs, such that the following two conditions are satisfied:

  (a) A run is in $\mathbf{Lr}^A_e$ iff all of its nonempty finite initial segments are in $\mathbf{Lr}^A_e$.

  (b) No run containing the (whatever-labeled) move ♠ is in $\mathbf{Lr}^A_e$.

  Elements of $\mathbf{Lr}^A_e$ are called **legal runs** of $A$ on input $e$, and all other runs called **illegal runs** on that input. In particular, if the last move of the shortest illegal-on-$e$ initial segment of $\Gamma$ is $\wp$-labeled, then $\Gamma$ is said to be a $\wp$-**illegal run** of $A$ on $e$.

- $\mathbf{Wn}^A$ is a function that sends every input $e$ and run $\Gamma$ to a player $\mathbf{Wn}^A_e \langle \Gamma \rangle$ ($= \top$ or $\bot$), such that the following condition is satisfied:

  (c) If $\Gamma$ is a $\wp$-illegal run of $A$ on $e$, then $\mathbf{Wn}^A_e \langle \Gamma \rangle \neq \wp$.

The intuitive meaning of the $\mathbf{Lr}$ component of a game is that it tells us what runs are legal; and the $\mathbf{Wn}$ component tells us who has won a given run of the game. What runs are legal or who the winner is may depend on what the input is, so both $\mathbf{Lr}$ and $\mathbf{Wn}$ take input $e$ as a parameter.

---
[3]What we call "input" in this paper is referred to as *valuation* in [7].

Understanding an *illegal move* as a move adding which (with the corresponding label) to the given position makes it illegal, condition (a) of the above definition corresponds to the intuition that a run is legal iff no illegal moves have been made in it, which automatically implies that the empty position $\langle\rangle$ is a legal run of every game. Condition (b) formalizes the status of ♠ as an "always-illegal" move. And condition (c) means that an illegal run is always lost by the player who has made the first illegal move. When modeling real-life interactive tasks, such as server-client or robot-environment interaction, illegal moves will usually mean "impossible actions". However, our approach — for generality, flexibility and convenience — does not formally rule out the possibility of making illegal moves.

$\Gamma$ is said to be a **unilegal** run (position if finite) of game $A$ iff, for every input $e$, $\Gamma \in \mathbf{Lr}^A_e$. We use $\mathbf{LR}^A$ to denote the set of all unilegal runs of $A$. A **unistructural** game is a game $A$ whose $\mathbf{Lr}$ component does not depend on input, i.e., for any two inputs $e_1$ and $e_2$, $\mathbf{Lr}^A_{e_1} = \mathbf{Lr}^A_{e_2}$. All examples of games discussed in this paper, as well as all games expressible in the language of logic **CL1** to which the paper is devoted, are unistructural. For such games, there is no difference between "legal" and "unilegal", and in appropriate contexts we will be using these two terms as synonyms.

Throughout this paper, unless otherwise specified, by a **predicate** we mean a set $P$ of infinite sequences of natural numbers, i.e. a collection of inputs, with "$P$ *is true at* $e$" or simply "$P(e)$ *is true*" meaning that $e \in P$, and "*false*" meaning "not true". Such a $P$ is said to be **finitary** iff there is an integer $k$ such that $P$ only depends on the first $k$ terms of the input — that is, for any two inputs $e_1$ and $e_2$ that (as sequences of numbers) agree on their first $k$ terms, we have $e_1 \in P$ iff $e_2 \in P$. Predicates that are not finitary are said to be **infinitary**. Of course, every finitary predicate that only depends on the first $k$ terms of the input can be thought of as a $k$-ary arithmetical relation, and vice versa. Say, the usual arithmetical relation $-_1 < -_2$ is the predicate that is true at $e = \langle a_1, a_2, a_3, \ldots \rangle$ iff $a_1 < a_2$.

We say that game $A$ is **elementary** iff, for every input $e$, $\mathbf{Lr}^A_e = \{\langle\rangle\}$. Thus, elementary games are games that have no legal moves: the empty run $\langle\rangle$ is the only legal run of such games. As this was noted in Section 2, we think of predicates as elementary games. In particular, a predicate can be understood as the (unique) elementary game $A$ such that $\mathbf{Wn}^A_e \langle\rangle = \top$ iff the predicate is true at $e$. And vice versa: every elementary game $A$ can be understood as the predicate that is true at $e$ iff $\mathbf{Wn}^A_e \langle\rangle = \top$. With this correspondence in mind, we will be using the terms "predicate" and "elementary game" as synonyms.

Now we formally define the main operations on games together with what we call *trivial games*, for which we use the same symbols $\top$ and $\bot$ as for the two players.

**Definition 3.2** In each of the following clauses, $\Phi$ ranges over *nonempty* positions, and $\Gamma$ ranges over legal runs of the game that is being defined. In view of Definition 3.1, it would be sufficient to define $\mathbf{Lr}$ and $\mathbf{Wn}$ only for this sort of $\Phi$ and $\Gamma$, respectively, for then $\mathbf{Lr}$ and $\mathbf{Wn}$ can be uniquely extended to all runs. $e$ ranges over all inputs. $A, A_1, \ldots, A_n$ ($n \geq 2$) are any games. The notation $\bar{\Phi}$ in clause 1 means the result of reversing (interchanging $\top$ with $\bot$) the labels in all labeled moves of $\Phi$. And the notation $\Phi^{i.}$ in clauses 2 and 3 means the result of removing from $\Phi$ all labeled moves except those of the form $\wp i.\alpha$ ($\wp \in \{\top, \bot\}$), and then deleting the prefix '$i.$' in the remaining moves, i.e. replacing each such $\wp i.\alpha$ by $\wp\alpha$. Similarly for $\bar{\Gamma}$, $\Gamma^{i.}$.

1. **Negation** $\neg A$:
    - $\Phi \in \mathbf{Lr}^{\neg A}_e$ iff $\bar{\Phi} \in \mathbf{Lr}^A_e$;
    - $\mathbf{Wn}^{\neg A}_e \langle\Gamma\rangle = \top$ iff $\mathbf{Wn}^A_e \langle\bar{\Gamma}\rangle = \bot$.

2. **Parallel conjunction** $A_1 \wedge \ldots \wedge A_n$:
    - $\Phi \in \mathbf{Lr}^{A_1 \wedge \ldots \wedge A_n}_e$ iff every move of $\Phi$ has the prefix '1.' or ... or '$n$.' and, for each $i \in \{1, \ldots, n\}$, $\Phi^{i.} \in \mathbf{Lr}^{A_i}_e$;
    - $\mathbf{Wn}^{A_1 \wedge \ldots \wedge A_n}_e \langle\Gamma\rangle = \top$ iff, for each $i \in \{1, \ldots, n\}$, $\mathbf{Wn}^{A_i}_e \langle\Gamma^{i.}\rangle = \top$.

3. **Parallel disjunction** $A_1 \vee \ldots \vee A_n$:
    - $\Phi \in \mathbf{Lr}^{A_1 \vee \ldots \vee A_n}_e$ iff every move of $\Phi$ has the prefix '1.' or ... or '$n$.' and, for each $i \in \{1, \ldots, n\}$, $\Phi^{i.} \in \mathbf{Lr}^{A_i}_e$;

- $\mathbf{Wn}_e^{A_1 \vee \ldots \vee A_n} \langle \Gamma \rangle = \bot$ iff, for each $i \in \{1, \ldots, n\}$, $\mathbf{Wn}_e^{A_i} \langle \Gamma^{i\cdot} \rangle = \bot$.

4. **Choice conjunction** $A_1 \sqcap \ldots \sqcap A_n$:
   - $\Phi \in \mathbf{Lr}_e^{A_1 \sqcap \ldots \sqcap A_n}$ iff $\Phi = \langle \bot i, \Psi \rangle$, where $i \in \{1, \ldots, n\}$ and $\Psi \in \mathbf{Lr}_e^{A_i}$;
   - $\mathbf{Wn}_e^{A_1 \sqcap \ldots \sqcap A_n} \langle \Gamma \rangle = \bot$ iff $\Gamma = \langle \bot i, \Delta \rangle$, where $i \in \{1, \ldots, n\}$ and $\mathbf{Wn}_e^{A_i} \langle \Delta \rangle = \bot$.

5. **Choice disjunction** $A_1 \sqcup \ldots \sqcup A_n$:
   - $\Phi \in \mathbf{Lr}_e^{A_1 \sqcup \ldots \sqcup A_n}$ iff $\Phi = \langle \top i, \Psi \rangle$, where $i \in \{1, \ldots, n\}$ and $\Psi \in \mathbf{Lr}_e^{A_i}$;
   - $\mathbf{Wn}_e^{A_1 \sqcup \ldots \sqcup A_n} \langle \Gamma \rangle = \top$ iff $\Gamma = \langle \top i, \Delta \rangle$, where $i \in \{1, \ldots, n\}$ and $\mathbf{Wn}_e^{A_i} \langle \Delta \rangle = \top$.

6. **Negative trivial game** $\bot$: $\quad \Phi \notin \mathbf{Lr}_e^\bot; \quad \mathbf{Wn}_e^\bot \langle \rangle = \bot$.

7. **Positive trivial game** $\top$: $\quad \Phi \notin \mathbf{Lr}_e^\top; \quad \mathbf{Wn}_e^\top \langle \rangle = \top$.

8. **Parallel implication**, or **reduction** $A_1 \to A_2$ is defined as $(\neg A_1) \vee A_2$.

Notice the perfect symmetry/duality between $\wedge$ and $\vee$, $\sqcap$ and $\sqcup$, $\bot$ and $\top$: the definition of each of these operations can be obtained from the definition of its dual by interchanging $\top$ with $\bot$. We earlier characterized legal plays over $A_1 \wedge A_2$ or $A_1 \vee A_2$ as plays "on two boards". According to the above definition, making a move $\alpha$ on "board" #$i$ is technically done by prefixing $\alpha$ with '$i.$'.

**Exercise 3.3** Verify that the following equalities hold for all games:

1. $A = \neg\neg A$;
2. $A \wedge B = \neg(\neg A \vee \neg B)$;
3. $A \sqcap B = \neg(\neg A \sqcup \neg B)$.

**Exercise 3.4** Let $e$ be any input and $A$ and $B$ any predicates that are both true at $e$. Verify that either of the runs $\langle \bot 1.1, \top 2.1.2 \rangle$, $\langle \top 2.1.2, \bot 1.1 \rangle$ is a unilegal and $\top$-won (on $e$) run of the game $(A \sqcup \neg A) \to ((\neg B \sqcup B) \wedge \top)$, i.e. — by Exercise 3.3 — of the game $(\neg A \sqcap A) \vee ((\neg B \sqcup B) \wedge \top)$. How about the runs $\langle \rangle$, $\langle \bot 1.1 \rangle$, $\langle \top 2.1.2 \rangle$?

The following fact, already observed in [7], is rather straightforward and hardly requires any proof:

**Lemma 3.5**
 a) All elementary games are unistructural.
 b) The operations $\neg, \wedge, \vee, \to, \sqcap, \sqcup$ preserve the unistructural property of games.

One more game operation that we are going to look at is that of *prefixation*, which is somewhat reminiscent of the modal operator(s) of dynamic logic. This operation takes two arguments: a game $A$ and a position $\Phi$ that must be a unilegal position of $A$ (otherwise the operation is undefined).

**Definition 3.6** Assume $\Phi$ is a unilegal position of a game $A$. The $\Phi$-**prefixation** of $A$, denoted $\langle \Phi \rangle A$, is defined as follows:

- $\mathbf{Lr}_e^{\langle \Phi \rangle A} = \{\Gamma \mid \langle \Phi, \Gamma \rangle \in \mathbf{Lr}_e^A\}$.
- $\mathbf{Wn}_e^{\langle \Phi \rangle A} \langle \Gamma \rangle = \mathbf{Wn}_e^A \langle \Phi, \Gamma \rangle$.

Intuitively, $\langle \Phi \rangle A$ is the game playing which means playing $A$ starting (continuing) from position $\Phi$. That is, $\langle \Phi \rangle A$ is the game to which $A$ evolves (will be "*brought down*") after the moves of $\Phi$ have been made. We have already used this intuition when explaining the meaning of choice operations: we said that after $\top$ makes an initial move $i \in \{1, 2\}$, the game $A_1 \sqcup A_2$ continues as $A_i$. What this meant was nothing but that $\langle \top i \rangle (A_1 \sqcup A_2) = A_i$. The following Proposition 3.7 summarizes this sort of a characterization of choice operations and extends it to other operations too. It tells us what the unilegal initial moves (i.e. one-move unilegal positions) of a given game $A$ are, and to what new game they bring $A$ down.

For a player $\wp$, we use the notation $\bar{\wp}$ to mean "the adversary of $\wp$", i.e. if $\wp = \top$, then $\bar{\wp} = \bot$, and if $\wp = \bot$, then $\bar{\wp} = \top$.

**Proposition 3.7** Below $A, B, A_1, \ldots, A_n$ ($n \geq 2$) are any games, $\alpha, \beta$ any moves, and $\wp$ either player; in each subclause (b), the position used in prefixation is a unilegal position of the game to which it is applied. We have:

1. (a) $\langle \wp\alpha \rangle \in \mathbf{LR}^{\neg A}$ iff $\langle \bar{\wp}\alpha \rangle \in \mathbf{LR}^{A}$;
   (b) $\langle \wp\alpha \rangle \neg A = \neg \langle \bar{\wp}\alpha \rangle A$.

2. (a) $\langle \wp\alpha \rangle \in \mathbf{LR}^{A_1 \wedge \ldots \wedge A_n}$ iff $\alpha = i.\beta$, where $i \in \{1, \ldots, n\}$ and $\langle \wp\beta \rangle \in \mathbf{LR}^{A_i}$;
   (b) $\langle \wp i.\beta \rangle (A_1 \wedge \ldots \wedge A_n) = A_1 \wedge \ldots \wedge A_{i-1} \wedge \langle \wp\beta \rangle A_i \wedge A_{i+1} \wedge \ldots \wedge A_n$.

3. (a) $\langle \wp\alpha \rangle \in \mathbf{LR}^{A_1 \vee \ldots \vee A_n}$ iff $\alpha = i.\beta$, where $i \in \{1, \ldots, n\}$ and $\langle \wp\beta \rangle \in \mathbf{LR}^{A_i}$;
   (b) $\langle \wp i.\beta \rangle (A_1 \vee \ldots \vee A_n) = A_1 \vee \ldots \vee A_{i-1} \vee \langle \wp\beta \rangle A_i \vee A_{i+1} \vee \ldots \vee A_n$.

4. (a) $\langle \wp\alpha \rangle \in \mathbf{LR}^{A \to B}$ iff $\alpha = i.\beta$, where $\begin{cases} i = 1 \text{ and } \langle \bar{\wp}\beta \rangle \in \mathbf{LR}^{A}, \text{ or} \\ i = 2 \text{ and } \langle \wp\beta \rangle \in \mathbf{LR}^{B}; \end{cases}$
   (b) $\begin{cases} \langle \wp 1.\beta \rangle (A \to B) = (\langle \bar{\wp}\beta \rangle A) \to B; \\ \langle \wp 2.\beta \rangle (A \to B) = A \to (\langle \wp\beta \rangle B). \end{cases}$

5. (a) $\langle \wp\alpha \rangle \in \mathbf{LR}^{A_1 \sqcap \ldots \sqcap A_n}$ iff $\wp = \bot$ and $\alpha = i \in \{1, \ldots, n\}$;
   (b) $\langle \bot i \rangle (A_1 \sqcap \ldots \sqcap A_n) = A_i$.

6. (a) $\langle \wp\alpha \rangle \in \mathbf{LR}^{A_1 \sqcup \ldots \sqcup A_n}$ iff $\wp = \top$ and $\alpha = i \in \{1, \ldots, n\}$;
   (b) $\langle \top i \rangle (A_1 \sqcup \ldots \sqcup A_n) = A_i$.

The above fact is known from [7]. Its proof consists in a routine analysis of the relevant definitions.

The language of logic **CL1** does not have any constructs corresponding to prefixation. However, this operation will be heavily exploited in our soundness and completeness proof for **CL1**. Generally, prefixation is very handy in visualizing a (unilegal) run of a given game $A$. In particular, every (sub)position $\Phi$ of such a run can be represented by, or thought of as, the game $\langle \Phi \rangle A$.

**Example 3.8** Remember game $G = (A \sqcup \neg A) \to \big((\neg B \sqcup B) \wedge \top\big)$ from Exercise 3.4. Based on Proposition 3.7, the run $\langle \top 2.1.2, \bot 1.1 \rangle$ is a unilegal run of $G$, and to it corresponds the following sequence of games:

(i) $(A \sqcup \neg A) \to \big((\neg B \sqcup B) \wedge \top\big)$, i.e. $G$, i.e. $\langle \rangle G$;

(ii) $(A \sqcup \neg A) \to (B \wedge \top)$, i.e. $\langle \top 2.1.2 \rangle$(i), i.e. $\langle \top 2.1.2 \rangle G$;

(iii) $A \to (B \wedge \top)$, i.e. $\langle \bot 1.1 \rangle$(ii), i.e. $\langle \top 2.1.2, \bot 1.1 \rangle G$.

The run hits the predicate $A \to (B \wedge \top)$ and, depending on whether the latter is true or false at a given input, the (present run of the) original game $G$ is won or lost by the machine on that input. If the run had stopped at (ii) or (i), it would be won by $\top$ because of the $\sqcup$-game in the antecedent.

There is an extensive literature on "game-style" models of computation in complexity theory: alternating machines, interactive proof systems, etc. An expert will notice that our approach, which is concerned with computability rather than complexity — just as the approaches to interaction taken in [4, 12] — is hardly related to that line of research, and the similarity is more terminological than conceptual. From the other, 'games in logic' or 'game semantics for linear logic' line, the closest to our present approach to games is Blass's [3] model, and less so the more recent studies by Abramsky, Jagadeesan, Hyland, Ong and others. This is not a survey paper, and we will only occasionally comment on how other approaches compare with ours. See [7] for more.

One of the main distinguishing features of our games among the other concepts of games studied in the logical literature (including the one tackled by the author [5] earlier) is the absence of what in [2] is called *procedural rules* — rules strictly regulating who and when should move, the most standard procedural rule being the one according to which players should take turns in alternating order. In our games, either player is free to make any move at any time. Such games can be called *free*, while games where in any given situation only one of the players is allowed to move called *strict*. Strict games can be thought of as special cases of our free games, where the **Lr** component is such that in any given position at most one of the players has legal moves (even though this would not formally preclude the "wrong" player from making a move, at least such a move would not go without penalty). Our games are thus most general of all two-player, two-outcome games. This makes them the most powerful and flexible modeling tool for interactive tasks. It also makes our definitions of game operations as simple, compact and natural as they could be, and allows us to adequately capture certain intended intuitions associated with those operations. Consider the game *Chess∧Chess*. Assume an agent plays this two-board game over the Internet against two independent adversaries that, together, form the (one) environment for the agent. Playing white on both boards, in the initial position of this game only the agent has legal moves. But once such a move is made, say, on the left board, the picture changes. Now both the agent and the environment have legal moves: the agent may make another opening move on the right board, while the environment — in particular, adversary #1 — may make a reply move on the left board. This is a situation where which player 'can move' is no longer strictly determined, so the next player to move will be the one who can or wants to act sooner. A strict-game approach would impose some additional conditions uniquely determining the next player to move. Such conditions would most likely be artificial and not quite adequate, for the situation we are trying to model is a concurrent play on two boards against two independent adversaries, and we cannot or should not expect any coordination between their actions. Most of the compound tasks that we perform in everyday life are free rather than strict, and so are most computer communication/interaction protocols. A strict understanding of ∧ would essentially mean some sort of an (in a sense interlaced but still) sequential rather than truly parallel/concurrent combination of tasks, where no steps in one component can be made until receiving a response in the other component, contrary to the very (utility-based) idea of parallel/distributed computation.

The class of games in the above-defined sense is general enough to model anything that we would call a (two-agent, two-outcome) interactive problem. However, it is too general. There are games where the chances of a player to succeed essentially depend on the relative speed at which its adversary responds, and we do not want to consider that sort of games meaningful computational problems. A simple example would be the game where all non-♠ moves are legal and that is won by the player who moves first. This is merely a contest of speed. Below we define a subclass of games called *static games*. Intuitively, they are games where speed is irrelevant: in order to succeed, only matters *what* to do (strategy) rather than *how fast* to do (speed). In particular, if a player can succeed when acting fast in such a game, it will remain equally successful acting the same way but slowly. This releases the player from any pressure for time and allows it to select its own pace for the game.

We say that run $\Delta$ is a $\wp$-**delay** of run $\Gamma$ iff the following two conditions are satisfied:

- for each player $\wp'$, the subsequence of $\wp'$-labeled moves of $\Delta$ is the same as that of $\Gamma$, and
- for any $n, k \geq 1$, if the $n$th $\wp$-labeled move is made later than (is to the right of) the $k$th $\bar{\wp}$-labeled move in $\Gamma$, then so is it in $\Delta$.

This means that in $\Delta$ each player has made the same sequence of moves as in $\Gamma$, only, in $\Delta$, $\wp$ might have been acting with some delay. Then we say that a game $A$ is **static** iff, whenever $\mathbf{Wn}_e^A\langle\Gamma\rangle = \wp$ and $\Delta$ is a $\wp$-delay of $\Gamma$, we have $\mathbf{Wn}_e^A\langle\Delta\rangle = \wp$.

Remember the game $G$ from Example 3.8, among the unilegal runs of which are $\Gamma = \langle \top 2.1.2, \bot 1.1 \rangle$ and $\Delta = \langle \bot 1.1, \top 2.1.2 \rangle$. In view of the following Proposition 3.9, $G$ is static. $\Delta$ is a $\top$-delay of $\Gamma$ and hence, whenever $\Gamma$ is won by $\top$, so should be $\Delta$. This is indeed so, because the two runs yield the same last game (iii) on which and only on which the outcome of the play (on a given input) depends.

The first clause of the following fact is a rather straightforward observation, and the second clause — also extended to other computability-logic operations — has been proven in [7], Theorem 14.1:

**Proposition 3.9**
  *a) All elementary games are static.*
  *b) The operations $\neg, \wedge, \vee, \rightarrow, \sqcap, \sqcup$ preserve the static property of games.*

One of the main theses on which computability logic relies philosophically is that the concept of static games is an adequate formal counterpart of our intuitive notion of "pure", speed-independent computational problems. See Section 4 of [7] for a detailed discussion and examples in support of this thesis.

Now we are ready to formally clarify what we mean by computational problems: we use the term "(interactive) **computational problem**" (or simply "**problem**") as a synonym of "static game".

## 4 Interactive computability

Now that we know what computational problems are, it is time to clarify what their computability means. The definitions given in this section are semiformal. All of the omitted technical details are however rather standard or irrelevant and can be easily restored by anyone familiar with Turing machines. If necessary, the corresponding detailed definitions can be found in Part II of [7].

As we remember, the central point of our philosophy is to require that agent $\top$ be implementable as a computer program, with effective and fully determined behavior. On the other hand, the behavior of agent $\bot$ can be arbitrary. This intuition is captured by the model of interactive computation where $\top$ is formalized as what we call HPM ("Hard-Play Machine").

Before we describe the HPM model, let us revisit ordinary Turing machines (TM). One of many equivalent versions of the TM model is the one that has three separate tapes: the read-only *input tape*, the read/write *work tape*, and the *output tape*. Among the reasons for wanting to have the three tapes separate might be that this model easily accommodates the possibility of infinite inputs and outputs. In particular, we now think of both input and output as (possibly infinite) *sequences* of finite strings rather than as individual (possibly infinite) strings. When such sequences are represented ("spelled") on tapes, some specially designated symbols are used as 'beginning of string' markers to separate strings. Specifically, we assume that symbol $\top$ is the beginning-of-string marker for the output tape. One may expect that the output tape be write-only. However, an equivalent version of ordinary TM model is the one where the output tape is instead read-only. Rather than directly printing a string $\top\alpha$ symbol-by-symbol on the output tape, the machine constructs $\alpha$ at the beginning of its work tape, delimits its end by positioning the read-write head there, and enters one of the specially designated states called *output states*.[4] This action results in $\top\alpha$ being automatically appended to the current contents of the output tape (which technically does not count as a write as there is no write head involved), and is thought of by us as making move $\alpha$ by the machine. Thus, the version of "ordinary" TM that we have just described is a model of non-interactive computation, with the machine being the only agent who can make moves and, this way, having full control over what output is being generated.

An **HPM** $\mathcal{H}$ is nothing but a TM in the above-described sense with one "little" difference: now it is not only the machine that can make moves, but also the environment. That is, at any time, an arbitrary new string $\bot\beta$ — with $\beta$ thought of as environment's move — can be nondeterministically appended to the contents of the output tape. At any given time the output tape thus spells the "current position" of the game, which is fully visible to the machine as it has read access to that tape. As always, the computation proceeds in discrete steps, also called *clock cycles*. From the above description it is clear that $\mathcal{H}$ can make at most one move per clock cycle. On the other hand, there are no limitations to the relative speed of the environment, so the latter can make any finite number of moves per cycle. This corresponds to the intuition that not only the strategy, but also the speed of the environment can be arbitrary. For technical clarity, we assume that the output tape remains stable during a clock cycle and is updated only on a transition from one cycle to another, and that if, during a given cycle, $\mathcal{H}$ makes a move $\alpha$ and the environment makes moves $\beta_1, \ldots, \beta_n$, then the position spelled on the output tape throughout the next cycle will be the result of appending $\langle \bot\beta_1, \ldots, \bot\beta_n, \top\alpha \rangle$ to the current position.

A **configuration** of $\mathcal{H}$ is defined in the standard way: this is a full description of the ("current") state of the machine, the locations of its three scanning heads, and the contents of its tapes, with the exception that, in order to make finite descriptions of configurations possible, we do not formally include a description of the unchanging (and possibly essentially infinite) contents of the input tape as a part of configuration. The *initial configuration* is the configuration where $\mathcal{H}$ is in its initial state and the work and output tapes are empty. A configuration $C'$ is said to be an *e*-**successor** of configuration $C$ if, when input $e$ is spelled on the input tape, $C'$ can legally follow $C$ in the standard sense, based on the transition function of the machine and accounting for the possibility of the above-described nondeterministic updates of the contents of the output tape. An *e*-**computation branch** of $\mathcal{H}$ is a sequence of configurations of $\mathcal{H}$ where the first configuration is the initial configuration and every other configuration is an $e$-successor of the previous one. Thus, the set of all $e$-computation branches captures all possible scenarios (on input $e$) corresponding to different behaviors by $\bot$.

---

[4]The terms used in [7, 8] for what we now call "output state", "input tape" and "output tape" are "move state", "valuation tape" and "run tape", respectively.

Each $e$-computation branch $B$ of $\mathcal{H}$ incrementally spells — in the obvious sense — some run $\Gamma$ on the output tape, which we call the **run spelled by** $B$. Then, for a game $A$, we write $\mathcal{H} \models_e A$ and say that $\mathcal{H}$ **wins** $A$ **on** $e$ iff, whenever $\Gamma$ is the run spelled by some $e$-computation branch of $\mathcal{H}$, $\mathbf{Wn}_e^A \langle \Gamma \rangle = \top$. We write $\mathcal{H} \models A$ and say that $\mathcal{H}$ **wins** (**computes**, **solves**) $A$ iff $\mathcal{H} \models_e A$ for every input $e$. Finally, we write $\models A$ and say that $A$ is **computable** iff there is an HPM $\mathcal{H}$ with $\mathcal{H} \models A$.

The HPM model of interactive computation seemingly strongly favors the environment in that the latter may be arbitrarily faster than the machine. What happens if we start limiting the speed of the environment? The answer is: *nothing* as far as computational problems, i.e. static games, are concerned. The model of computation that we call EPM ("Easy-Play Machine") takes the idea if limiting the speed of the environment to the extreme by giving the machine full control over when the environment can move and when it should wait; yet, as it turns out, the EPM model yields the same class of computable problems as the HPM model does. EPMs will be discussed later in Section 6.

## 5 Logic CL1

The language of logic **CL1** is obtained by augmenting the language of classical propositional logic with two additional operators $\sqcap$ and $\sqcup$. This language has two **logical atoms**: $\top$ and $\bot$, and infinitely many **non-logical atoms** called "propositional letters" in classical logic. Throughout the rest of this paper by a **formula** we mean a formula of this language. The definition is standard: the set of formulas is the smallest set of expressions such that atoms are formulas and, as long as $F_1, \ldots, F_n$ ($n \geq 2$) are formulas, so are $\neg(F_1)$, $(F_1) \to (F_2)$, $(F_1) \wedge \ldots \wedge (F_n)$, $(F_1) \vee \ldots \vee (F_n)$, $(F_1) \sqcap \ldots \sqcap (F_n)$, $(F_1) \sqcup \ldots \sqcup (F_n)$. Parentheses will often be omitted in accordance with standard conventions.

An **interpretation** is a function $^*$ that sends each non-logical atom $p$ to an elementary game $p^*$. This mapping extends to all formulas by commuting with each operator. That is: $\bot^* = \bot$; $\top^* = \top$; $(\neg F)^* = \neg(F^*)$; $(F_1 \to F_2)^* = F_1^* \to F_2^*$; $(F_1 \wedge \ldots \wedge F_n)^* = F_1^* \wedge \ldots \wedge F_n^*$; and similarly for $\vee, \sqcap, \sqcup$. For a formula $F$, we will say "$^*$ **interprets** $F$ **as** $A$" to mean that $F^* = A$.

**Proposition 5.1** *For any formula $F$ and interpretation $^*$, the game $F^*$ is unistructural and static.*

**Proof.** Immediately from Propositions 3.5, 3.9 and the fact that, by definition, $F^*$ is a $\neg, \wedge, \vee, \to, \sqcap, \sqcup$-combination of elementary games. $\square$

**Definition 5.2** A formula $F$ is said to be **valid** iff, for every interpretation $^*$, $\models F^*$.

To axiomatize the set of valid formulas, we need some technical preliminaries. Understanding $F \to G$ as an abbreviation for $\neg F \vee G$, a **positive** (resp. **negative**) **occurrence** of a subformula is one that is in the scope of an even (resp. odd) number of occurrences of $\neg$. A **surface occurrence** of a subformula is an occurrence that is not in the scope of $\sqcap$ or $\sqcup$. A formula not containing $\sqcap, \sqcup$ is said to be **elementary**. The **elementarization** of formula $F$ is the result of replacing in $F$ all subformulas of the form $G_1 \sqcup \ldots \sqcup G_n$ by $\bot$ and all subformulas of the form $G_1 \sqcap \ldots \sqcap G_n$ by $\top$. A formula is said to be **stable** iff its elementarization is classically valid. Otherwise it is **instable**.

**Definition 5.3** Logic **CL1** is given by the following two rules:

(a) $\vec{H} \vdash F$, where $F$ is stable and $\vec{H}$ is the smallest set of formulas such that, whenever $F$ has a positive (resp. negative) surface occurrence of a subformula $G_1 \sqcap \ldots \sqcap G_n$ (resp. $G_1 \sqcup \ldots \sqcup G_n$), for each $i \in \{1, \ldots, n\}$, $\vec{H}$ contains the result of replacing this occurrence in $F$ by $G_i$.

(b) $H \vdash F$, where $H$ is the result of replacing in $F$ a negative (resp. positive) surface occurrence of a subformula $G_1 \sqcap \ldots \sqcap G_n$ (resp. $G_1 \sqcup \ldots \sqcup G_n$) by $G_i$ for some $i \in \{1, \ldots, n\}$.

Axioms are not explicitly stated, but note that the set $\vec{H}$ of premises of Rule **(a)** can be empty, in which case the conclusion $F$ of that rule acts as an axiom. Even though this may not be immediately obvious, **CL1** essentially *is* a (refined sort of) Gentzen-style system. Consider, for example, Rule **(b)**. It is very similar to the $\sqcup$-introduction rule of linear logic. The only difference is that while linear or affine logics require that $G_i$ be a $\vee$-disjunct of the premise, **CL1** (explicitly) allows it to be any positive surface occurrence. This is called *deep inference* in the calculus of structures.

In Examples 5.4, 5.5 and Exercise 5.6 below, $p, q, r, s$ are pairwise distinct non-logical atoms.

**Example 5.4** Let us see that
$$\mathbf{CL1} \vdash \big((p \to q) \sqcap (p \to r)\big) \to \big(p \to (q \sqcap r)\big).$$

1. $(p \to q) \to (p \to q)$  (from $\emptyset$ by Rule **(a)**)
2. $\big((p \to q) \sqcap (p \to r)\big) \to (p \to q)$ (from 1 by Rule **(b)**)
3. $(p \to r) \to (p \to r)$  (from $\emptyset$ by Rule **(a)**)
4. $\big((p \to q) \sqcap (p \to r)\big) \to (p \to r)$ (from 3 by Rule **(b)**)
5. $\big((p \to q) \sqcap (p \to r)\big) \to \big(p \to (q \sqcap r)\big)$ (from {2,4} by Rule **(a)**)

**Example 5.5** On the other hand, we have
$$\mathbf{CL1} \not\vdash \big((p \to q) \sqcap (p \to r)\big) \to \big(p \to (q \wedge r)\big).$$

Indeed, the above formula is instable, so it could only be derived by Rule **(b)**. The premise of this rule should be either $(p \to q) \to \big(p \to (q \wedge r)\big)$ or $(p \to r) \to \big(p \to (q \wedge r)\big)$. In either case we deal with a formula that can be derived neither by Rule **(a)** (because the formula is instable) nor by Rule **(b)** (because the formula does not contain $\sqcap, \sqcup$).

**Exercise 5.6** Verify that:

1. $\mathbf{CL1} \vdash \big((p \sqcap q) \wedge (p \sqcap q)\big) \to (p \sqcap q)$;
2. $\mathbf{CL1} \not\vdash (p \sqcap q) \to \big((p \sqcap q) \wedge (p \sqcap q)\big)$;
3. $\mathbf{CL1} \not\vdash \Big(\big(p \wedge (r \sqcap s)\big) \sqcap \big(q \wedge (r \sqcap s)\big)\Big) \to \big((p \sqcap q) \wedge (r \sqcap s)\big)$.

As we probably just had a chance to notice, if $F$ is an elementary formula, then the only way to prove $F$ in **CL1** is to derive it by Rule **(a)** from the empty set of premises. In particular, this rule will be applicable when $F$ is stable, which for an elementary $F$ means nothing but that $F$ is a classical tautology. And vice versa: every classically valid formula is an elementary formula derivable in **CL1** by Rule **(a)** from the empty set of premises. Thus we have:

**Proposition 5.7** *The $\sqcap, \sqcup$-free fragment of **CL1** is exactly classical propositional logic.*

This is what we should have expected for, as noted in Section 2, when restricted to elementary problems — and elementary formulas are exactly the ones that represent such problems — the meanings of $\neg, \wedge, \vee, \to$ are classical. It is also rather obvious that **CL1** is decidable in at most polynomial space.

Below comes our main theorem, according to which **CL1** precisely describes the set of all valid principles of computability expressible in its language. This theorem is just a combination of Lemmas 7.1 and 8.2 proven later.

**Theorem 5.8** $\mathbf{CL1} \vdash F$ *iff $F$ is valid (any formula $F$). Furthermore:*

*a) There is an effective procedure that takes a **CL1**-proof of an arbitrary formula $F$ and constructs an HPM $\mathcal{H}$ such that, for every interpretation $*$, $\mathcal{H}$ computes $F^*$.*

*b) If $\mathbf{CL1} \not\vdash F$, then $F^*$ is not computable for some interpretation $*$ that interprets all atoms as finitary predicates.*

**CL1** is a fragment of the logic **FD** introduced in [7]. The language of **FD** is more expressive in that it has (two groups of) quantifiers as well as an additional sort of atoms called *general atoms*. Unlike our present type of atoms (called *elementary atoms* in [7]) that can only be interpreted as elementary games, general non-logical atoms can be interpreted as any computational problems. **CL1** is obtained from **FD** by mechanically deleting every rule but the propositional versions of the **A** and **B** groups of rules. Those deleted rules introduce general atoms or quantifiers that are alien to the language we now consider. Once a general atom or a quantifier is introduced, it never disappears in any later formulas of an **FD**-proof. Based on this observation, a formula in our present sense is provable in **FD** iff it is provable in **CL1**, so that **FD** is a conservative extension of **CL1**. It was conjectured in [7] (Conjecture 25.4) that **FD** is sound and complete with respect to computability semantics. Our Theorem 5.8 signifies a positive verification of that conjecture restricted to the language of **CL1**. There were two other major conjectures stated in [7]: Conjecture 24.4 and Conjecture 26.2. A positive verification of those two conjectures restricted to the language of **CL1** is also among the immediate consequences of our Theorem 5.8.

When restricted to the language of **CL1**, Conjecture 24.4 of [7] sounds as follows:

> *If a formula $F$ is not valid, then $F^*$ is not computable for some interpretation $^*$ that interprets all atoms as finitary predicates.*[5]

The significance of this conjecture is related to the fact that showing non-validity of a given formula by appealing to interpretations that interpret atoms as infinitary predicates generally could dramatically weaken such a non-validity statement. E.g., if the predicate $p^*$ is infinitary, then $p^* \sqcup \neg p^*$ may be incomputable just due to the fact that the machine can never finish reading all the relevant information from the input tape necessary to determine whether $p^*$ is true or false. On the other hand, once we restrict our considerations only to interpretations that interpret atoms as finitary predicates, the non-validity statement for $p \sqcup \neg p$ is indeed highly informative: the failure to solve $p^* \sqcup \neg p^*$ in such a case signifies fundamental limitations of algorithmic methods rather than just impossibility to obtain all the necessary external information. A positive solution to Conjecture 24.4 of [7] restricted to the language of **CL1** is implied by clause (b) of our Theorem 5.8 in conjunction with its soundness statement for **CL1**.

As for Conjecture 26.2, it was about equivalence between validity and another version of this notion called *uniform validity*. If we disabbreviate "$\models F^*$" as "$\exists \mathcal{H} \, (\mathcal{H} \models F^*)$", validity of $F$ in the sense of Definition 5.2 can be written as "$\forall \, ^* \, \exists \, \mathcal{H} \, (\mathcal{H} \models F^*)$". Reversing the order of quantification yields the following stronger property of uniform validity:

**Definition 5.9** A formula $F$ is said to be **uniformly valid** iff there is an HPM $\mathcal{H}$ such that, for every interpretation $^*$, $\mathcal{H} \models F^*$.

Intuitively, uniform validity means existence of an interpretation-independent solution: since no information regarding interpretation $^*$ comes as a part of input, the above HPM $\mathcal{H}$ with $\forall^*(\mathcal{H} \models F^*)$ will have to play in some standard, uniform way that would be successful for any possible $^*$.

The term "uniform" is borrowed from [1] as this understanding of validity in its spirit is close to that in Abramsky and Jagadeesan's tradition. The concepts of validity in Lorenzen's [9] tradition, or in the sense of Japaridze [6], also belong to this category. Common to those uniform-validity-style notions is that validity there is not defined as being "always true" (true=winnable) as this is the case with the classical understanding of this concept; in those approaches the concept of truth is often simply absent, and validity is treated as a basic concept in its own rights. As for (simply) validity, it is closer to validities in the sense of Blass [3] or Japaridze [5], and presents a direct generalization of the corresponding classical concept in that it indeed means being "true" (computable) in every particular setting.

Which of our two versions of validity is more interesting depends on the motivational standpoint. It is validity rather than uniform validity that tells us what can be computed in principle. So, a computability-theoretician would focus on validity. Mathematically, non-validity is generally by an order of magnitude more informative — and correspondingly harder to prove — than non-uniform-validity. Say, the non-validity of $p \sqcup \neg p$, with the above-quoted and now successfully verified Conjecture 24.4 of [7] in mind, means existence of solvable-in-principle yet algorithmically unsolvable problems — the fact that became known to the mankind only as late as in the 20th century. As for the non-uniform-validity of $p \sqcup \neg p$, it is trivial: of course there is no way to choose one of the two disjuncts that would be true for all possible values of $p$ because — as the stone age intellectuals were probably aware — some $p$ are true and some are false.

---

[5]The original formulation of Conjecture 24.4 imposes additional restrictions on interpretations. But those restrictions are automatically satisfied as long as we deal with the language of **CL1**.

On the other hand, it is uniform validity rather than validity that is of interest in more applied areas of computer science such as knowledgebase systems or resourcebase and planning systems (see Section 26 of [7] or Section 8 of [8]). In such applications we want a logic on which a universal problem-solving machine can be based. Such a machine would or should be able to solve problems represented by logical formulas without any specific knowledge of the meaning of their atoms, i.e. without knowledge of the actual interpretation. Remembering what was said about the intuitive meaning of uniform validity, this concept is exactly what fits the bill.

Anyway, the good news, signifying a successful verification of Conjecture 26.2 of [7] restricted to the language of **CL1**, is that the two concepts of validity yield the same logic. If $F$ is uniformly valid, then it is automatically also valid, as uniform validity is stronger than validity. Suppose now $F$ is valid. Then, by the completeness part of Theorem 5.8, $\mathbf{CL1} \vdash F$. But then, by the 'Furthermore' clause (a) of the same theorem, $F$ is uniformly valid. Thus, where — in accordance to our present convention — "formula" means formula of the language of **CL1**, we have:

**Theorem 5.10** *A formula is valid if and only if it is uniformly valid.*

## 6 Preliminaries for the soundness/completeness proof

The rest of this paper is almost exclusively devoted to a proof of our main Theorem 5.8. The present section contains some essential preliminaries. We start with a semiformal introduction of the alternative model of interactive computation called EPM. If necessary, detailed formal definitions can be found in [7].

An **EPM** is a machine defined in the same way as HPM, with the only difference that now the environment can (but is not obligated to) make a move only when the machine explicitly allows it to do so, the event that we call **granting permission**. Technically permission is granted by entering one of the specially designated states called **permission states**. The only requirement that the machine is expected to satisfy is that it should grant permission every once in a while; how long that "while" lasts, however, is totally up to the machine. This amounts to having full control over the speed of the adversary.

The above intuition is formalized as follows. After correspondingly redefining the '$e$-successor' relation — in particular, accounting for the condition that now a (one single) $\bot$-labeled move may be appended to the contents of the output tape only when the machine is in a permission state — the concepts of $e$-**computation branch** of an EPM, the run **spelled** by such a branch, etc. are defined in exactly the same way as for HPM. We say that a computation branch $B$ of an EPM is **fair** iff permission is granted infinitely many times in $B$. A **fair EPM** is an EPM whose every $e$-computation branch (for every input $e$) is fair. For a fair EPM $\mathcal{E}$ and input $e$ — just as in the case of HPMs — we write $\mathcal{E} \models_e A$ ("$\mathcal{E}$ **wins** $A$ **on** $e$") iff, whenever $\Gamma$ is the run spelled by some $e$-computation branch of $\mathcal{E}$, $\mathbf{Wn}_e^A \langle \Gamma \rangle = \top$; and $\mathcal{E} \models A$ ("$\mathcal{E}$ **wins** $A$", or "$\mathcal{E}$ **computes** $A$", or "$\mathcal{E}$ **solves** $A$") means that $\mathcal{E} \models_e A$ for every input $e$.

The following fact, proven in [7] (Theorem 17.2), establishes equivalence between the two models of computation for static games:

**Proposition 6.1** *Every fair EPM $\mathcal{E}$ can be effectively converted[6] into an HPM $\mathcal{H}$ such that $\mathcal{H}$ wins every static game that $\mathcal{E}$ does. And vice versa: every HPM $\mathcal{H}$ can be effectively converted into a fair EPM $\mathcal{E}$ such that $\mathcal{E}$ wins every static game that $\mathcal{H}$ does.*

The philosophical significance of this theorem is that it reinforces the thesis according to which static games are games that allow us to make full abstraction from speed. Its technical importance is related to the fact that the EPM-model is much more convenient when it comes to describing interactive algorithms/strategies, as we will have a chance to see later. The two models also act as natural complements to each other: we can meaningfully talk about the (uniquely determined) play between a given HPM and a given EPM, while this is impossible if both players are HPMs or both are EPMs. This fact will be essentially exploited in our completeness proof for **CL1**, where we describe environment's strategy as an EPM and show that no HPM can win the given game against such an EPM.

---

[6] Meaning that there is an effective function of the type {EPMs}→{HPMs} that successfully does the conversion as long as the EPM to which it is applied is fair.

Remember the notation $\bar{\Gamma}$ used in Definition 3.2 which means the result of reversing all labels in $\Gamma$. For a run $\Gamma$ and a computation branch $B$ of an HPM or EPM, we say that $B$ **cospells** $\Gamma$ iff $B$ spells $\bar{\Gamma}$. Intuitively, when machine $\mathcal{M}$ plays as $\bot$ (rather than $\top$), then the run that is generated by a given computation branch $B$ of $\mathcal{M}$ is the run cospelled (rather than spelled) by $B$, for the moves that $\mathcal{M}$ makes get the label $\bot$, and the moves that its adversary makes get the label $\top$.

**Lemma 6.2** *Assume $\mathcal{E}$ is a fair EPM, $\mathcal{H}$ any HPM, and $e$ any input. There are a uniquely defined e-computation branch $B_\mathcal{E}$ of $\mathcal{E}$ and a uniquely defined e-computation branch $B_\mathcal{H}$ of $\mathcal{H}$ — which we respectively call* **the $(\mathcal{E}, e, \mathcal{H})$-branch** *and* **the $(\mathcal{H}, e, \mathcal{E})$-branch** *— such that the run spelled by $B_\mathcal{H}$, called* **the $\mathcal{H}$ vs. $\mathcal{E}$ run on** *$e$, is the run cospelled by $B_\mathcal{E}$.*

When $\mathcal{H}, \mathcal{E}, e$ are as above, $\Gamma$ is the $\mathcal{H}$ vs. $\mathcal{E}$ run on $e$ and $A$ is a game with $\mathbf{Wn}_e^A\langle\Gamma\rangle = \top$ (resp. $\mathbf{Wn}_e^A\langle\Gamma\rangle = \bot$), we say that $\mathcal{H}$ **wins** (resp. **loses**) $A$ **against** $\mathcal{E}$ **on** $e$.

A strict proof of the above lemma can be found in [7] (Lemma 20.4), and we will not reproduce the formal proof here. Instead, the following intuitive explanation would suffice:

**Proof idea.** Assume $\mathcal{H}, \mathcal{E}, e$ are as in Lemma 6.2. The play that we are going to describe is the unique play generated when the two machines play against each other on input $e$, with $\mathcal{H}$ in the role of $\top$ and $\mathcal{E}$ in the role of $\bot$. We can visualize this play as follows. Most of the time during the play $\mathcal{H}$ remains inactive (sleeping); it is woken up only when $\mathcal{E}$ enters a permission state, on which event $\mathcal{H}$ makes a (one single) transition to its next computation step — that may or may not result in making a move — and goes back to sleep that will continue until $\mathcal{E}$ enters a permission state again, and so on. From $\mathcal{E}$'s perspective, $\mathcal{H}$ acts as a patient adversary who makes one or zero move only when granted permission, just as the EPM-model assumes. And from $\mathcal{H}$'s perspective, who, like a person under global anesthesia, has no sense of time during its sleep and hence can think that the wake-up events that it calls the beginning of a clock cycle happen at a constant rate, $\mathcal{E}$ acts as an adversary who can make any finite number of moves during a clock cycle (i.e. while $\mathcal{H}$ was sleeping), just as the HPM-model assumes. This scenario uniquely determines an e-computation branch $B_\mathcal{E}$ of $\mathcal{E}$ that we call the $(\mathcal{E}, e, \mathcal{H})$-branch, and an e-computation branch $B_\mathcal{H}$ of $\mathcal{H}$ that we call the $(\mathcal{H}, e, \mathcal{E})$-branch. What we call the $\mathcal{H}$ vs. $\mathcal{E}$ run on $e$ is the run generated in this play. In particular — since we let $\mathcal{H}$ play in the role of $\top$ — this is the run spelled by $B_\mathcal{H}$. $\mathcal{E}$, who plays in the role of $\bot$, sees the same run, only it sees the labels of that run in negative colors. That is, $B_\mathcal{E}$ cospells rather than spells that run. This is exactly what Lemma 6.2 asserts.

Now back to the language of **CL1**. By Proposition 5.1, every game $F^*$ expressible in it is unistructural. This allows us to relax our terminology and simply say "legal run of $F^*$" without specifying input, which should be understood as saying "unilegal run of $F^*$". Similarly for "$\wp$-illegal", where the input parameter can be safely omitted. Another helpful observation to make here is that the set $\mathbf{LR}^{F^*}$ of (uni)legal runs of $F^*$ is fully determined by $F$ and in no way does it depend on interpretation $^*$.

We will be using the notation

$$\|L\|$$

for the elementarization of formula $L$. The following lemma can be verified by a straightforward induction on the complexity of $L$:

**Lemma 6.3** $\mathbf{Wn}_e^{L^*}\langle\rangle = \top$ *iff the predicate $\|L\|^*$ is true at $e$ (any formula $L$, interpretation $^*$ and input $e$).*

Here we define a function that, for a formula $F$ and a surface occurrence $O$ in $F$, returns a string $\alpha$ called the *$F$-specification* of $O$, which is said to $F$-*specify* $O$. In particular:

- The occurrence of $F$ in itself is $F$-specified by the empty string.

- If $F = \neg G$, then an occurrence that happens to be in $G$ is $F$-specified by the same string that $G$-specifies that occurrence.

- If $F$ is $G_1 \wedge \ldots \wedge G_n$, $G_1 \vee \ldots \vee G_n$ or $G_1 \to G_2$, then an occurrence that happens to be in $G_i$ is $F$-specified by $i.\alpha$, where $\alpha$ is the $G_i$-specification of that occurrence.

Example: The second occurrence of $P \sqcup Q$ in $F = R \vee (P \sqcup Q) \vee \neg(P \to (R \wedge (P \sqcup Q)))$ is $F$-specified by the string '3.2.2.'.

Based on Proposition 3.7, the following lemma can be easily verified by induction on the complexity of $F$, the routine details of which we omit:

**Lemma 6.4** *For every formula $F$, move $\alpha$ and interpretation $*$ we have:*

*a) $\langle \bot\alpha \rangle \in \mathbf{LR}^{F^*}$ iff $\alpha = \beta i$, where $\beta$ is the $F$-specification of a positive (resp. negative) surface occurrence of a subformula $G_1 \sqcap \ldots \sqcap G_n$ (resp. $G_1 \sqcup \ldots \sqcup G_n$) and $i \in \{1, \ldots, n\}$. In this case $\langle \bot\alpha \rangle F^* = H^*$, where $H$ is the result of substituting in $F$ the above occurrence by $G_i$.*

*b) $\langle \top\alpha \rangle \in \mathbf{LR}^{F^*}$ iff $\alpha = \beta i$, where $\beta$ is the $F$-specification of a negative (resp. positive) surface occurrence of a subformula $G_1 \sqcap \ldots \sqcap G_n$ (resp. $G_1 \sqcup \ldots \sqcup G_n$) and $i \in \{1, \ldots, n\}$. In this case $\langle \top\alpha \rangle F^* = H^*$, where $H$ is the result of substituting in $F$ the above occurrence by $G_i$.*

# 7 Soundness of CL1

**Lemma 7.1** *If $\mathbf{CL1} \vdash F$, then $F$ is valid (any formula $F$).*

*Moreover, there is an effective procedure that takes a $\mathbf{CL1}$-proof of an arbitrary formula $F$ and constructs an HPM $\mathcal{H}$ such that, for every interpretation $*$, $\mathcal{H} \models F^*$.*

**Proof.** In view of Propositions 5.1 and 6.1, the present lemma — in particular, its 'Moreover' clause — is an immediate corollary of the following Lemma 7.2. □

**Lemma 7.2** *There is an effective procedure that takes a $\mathbf{CL1}$-proof of an arbitrary formula $F$ and constructs a fair EPM $\mathcal{E}$ such that, for every interpretation $*$, $\mathcal{E} \models F^*$.*

**Proof idea.** Every $\mathbf{CL1}$-proof of $F$, in fact, very directly encodes an input- and interpretation-independent winning EPM-strategy for $F^*$. Such a strategy/EPM acts depending on whether $F$ is obtained by Rule **(a)** or Rule **(b)** from its premise(s).

If $F$ is derived by Rule **(b)** from premise $H$, then, in view of Lemma 6.4b, there is a (legal) move $\alpha$ by the machine that brings the game down to $H$, in the sense that $\langle \top\alpha \rangle F^* = H^*$. E.g., if $F = (G_1 \sqcup G_2) \wedge (G_3 \sqcap G_4)$ and $H = G_2 \wedge (G_3 \sqcap G_4)$, then '1.2' is such a move. So, in order to win $F^*$, it is sufficient to make the move $\alpha$ and then win $H^*$. This is exactly what our EPM does: since the proof of $H$ is shorter than that of $F$, by the induction hypothesis, the EPM knows how to win $H^*$; so, it makes the move $\alpha$ and then switches to its winning strategy for $H^*$.

And if $F$ is derived by Rule **(a)**, then the machine chooses not to make any moves but rather to let the environment move. If the environment makes a move $\alpha$ — which must be legal for otherwise $\bot$ is immediately disqualified — then, for similar reasons as in the previous case, the resulting game $\langle \bot\alpha \rangle F^*$ will be nothing but $H^*$ for some premise $H$ of $F$. So, in this case the machine switches to its successful strategy for $H^*$ and eventually wins. And if the environment never makes a move, then the machine automatically wins. This is so because $F$, as a conclusion of Rule **(a)**, is stable, and the empty run of a game represented by a stable formula, in view of Lemma 6.3, is always won by $\top$.

**Proof.** Assume $\mathbf{CL1} \vdash F$. Let us fix a particular $\mathbf{CL1}$-proof of $F$. We will be referring to at as "the proof", and referring to formulas occurring in the proof as "proof formulas". We assume that the proof is a sequence (rather than tree) of formulas without repetitions or other redundancies, where each formula comes with a fixed *justification* — an indication of from what premises and by which rule it was derived. Based on the proof, we (effectively) construct the EPM $\mathcal{E}$ whose work is described as follows. At the beginning, this machine creates a record $E$ to hold proof formulas, initializes it to $F$, and then follows the following interactive algorithm:

**Procedure** LOOP: As long as $E$ is a proof formula, act depending on which of the two rules was used in the proof to derive $E$ from its premises:

**Case of Rule (a):** Keep granting permission until the adversary makes a move $\alpha$, then act depending on which of the following two subcases holds:

*Subcase (i):* $\alpha = \beta i$, where $\beta$ $E$-specifies a positive (resp. negative) surface occurrence of a subformula $G_1 \sqcap \ldots \sqcap G_n$ (resp. $G_1 \sqcup \ldots \sqcup G_n$) and $i \in \{1, \ldots, n\}$. Let $H$ be the result of substituting the above occurrence by $G_i$ in $E$. Then update (the content of) $E$ to $H$, and repeat LOOP.

*Subcase (ii):* $\alpha$ does not satisfy the condition of Subcase (i). Then go to an infinite loop in a permission state.

**Case of Rule (b):** Let $H$ be the premise of $E$ in the proof. $H$ is the result of substituting, in $E$, a certain negative (resp. positive) surface occurrence of a subformula $G_1 \sqcap \ldots \sqcap G_n$ (resp. $G_1 \sqcup \ldots \sqcup G_n$) by $G_i$ for some $i \in \{1, \ldots, n\}$. Let $\alpha$ be the $E$-specification of that occurrence. Then make the move $\alpha i$, update (the content of) $E$ to $H$, and repeat LOOP.

Note that the above description has no mention of input or interpretation, so $\mathcal{E}$ and its behavior remain the same no matter what those two parameters are.

Fix an arbitrary interpretation $^*$, an arbitrary input $e$ and an arbitrary $e$-computation branch $B$ of $\mathcal{E}$. Let $\Gamma$ be the run spelled by $B$. Our goal is to show that $B$ is fair and that $\mathbf{Wn}_e^{F^*}\langle \Gamma \rangle = \top$. Consider the work of $\mathcal{E}$ along $B$. Let $E_i$ denote the value of the record $E$ at the beginning of the $i$th iteration of LOOP. Evidently $E_{i+1}$ is always one of the premises of $E_i$ in the proof, which means that the 'As long as' condition of LOOP is always satisfied, and that LOOP is iterated only a finite number of times. Fix $l$ as the number of iterations of LOOP, and let $L = E_l$. The $l$th iteration deals with the case of Rule **(a)** — and, besides, *not* with Subcase (i) of that case — for otherwise there would be a next iteration. This guarantees that $\mathcal{E}$ will grant permission infinitely many times during the $l$th iteration, so that branch $B$ is indeed fair. Since $e$ was selected arbitrarily, we conclude that $\mathcal{E}$ is a fair EPM. And the fact that $L$ is derived by Rule **(a)** implies that

$$L \text{ is a stable proof formula.} \tag{1}$$

For each $i$ with $1 \leq i \leq l$, let $\Theta_i$ be the sequence of the (correspondingly labeled) moves made by the players by the beginning of the $i$th iteration of LOOP.

$$\text{For each } 1 \leq i \leq l, \ \Theta_i \in \mathbf{LR}^{F^*} \text{ and } \langle \Theta_i \rangle F^* = E_i^*. \tag{2}$$

This statement can be proven by induction on $i$. The basis case with $i = 1$ is trivial. Now consider an arbitrary $i$ with $1 \leq i < l$. By the induction hypothesis, $\Theta_i \in \mathbf{LR}^{F^*}$ and $\langle \Theta_i \rangle F^* = E_i^*$. If the $i$th iteration of LOOP deals with the case of Rule **(b)**, then exactly one move $\alpha$ is made during that iteration, and this move is by the machine, so that $\Theta_{i+1} = \langle \Theta_i, \top\alpha \rangle$. In view of Lemma 6.4b, $\langle \top\alpha \rangle \in \mathbf{LR}^{E_i^*}$ and $\langle \top\alpha \rangle E_i^* = E_{i+1}^*$. With the equalities $\Theta_{i+1} = \langle \Theta_i, \top\alpha \rangle$ and $E_i^* = \langle \Theta_i \rangle F^*$ in mind, the former then implies $\Theta_{i+1} \in \mathbf{LR}^{F^*}$ and the latter implies $\langle \Theta_{i+1} \rangle F^* = E_{i+1}^*$. Suppose now the $i$th iteration of LOOP deals with the case of Rule **(a)**. Then the machine does not make a move; its adversary must make a move $\alpha$, for otherwise $i$ would be the last iteration of LOOP, contrary to our assumption that $i < l$. Furthermore, for the same reason, this move $\alpha$ should satisfy the condition of Subcase (i). Then, arguing as in the previous case (only using Lemma 6.4a instead of 6.4b), we can again conclude that $\Theta_{i+1} \in \mathbf{LR}^{F^*}$ and $\langle \Theta_{i+1} \rangle F^* = E_{i+1}^*$.

$$\text{Either } \Gamma \text{ is a } \bot\text{-illegal run of } F^*, \text{ or } \mathbf{Wn}_e^{F^*}\langle \Gamma \rangle = \mathbf{Wn}_e^{L^*}\langle \rangle. \tag{3}$$

To prove this statement, suppose $\Gamma$ is not a $\bot$-illegal run of $F^*$. According to (2), $\langle \Theta_l \rangle F^* = L^*$, which implies that $\mathbf{Wn}_e^{F^*}\langle \Theta_l \rangle = \mathbf{Wn}_e^{L^*}\langle \rangle$. So, it would be sufficient to show that $\Theta_l = \Gamma$. But indeed, as the $l$th iteration of LOOP deals with the case of Rule **(a)**, during that iteration $\top$ does not make any moves; if $\bot$ makes a move $\alpha$ — since $l$ is the last iteration, — we deal with Subcase (ii), and then it is clear from Lemma 6.4a that $\langle \bot\alpha \rangle \notin \mathbf{LR}^{L^*}$ which, taking into account that (by (2)) $\Theta_l \in \mathbf{LR}^{F^*}$ and $L^* = \langle \Theta_l \rangle F^*$, implies that $\langle \Theta_l, \bot\alpha \rangle$ is a $\bot$-illegal position of $F^*$; but $\langle \Theta_l, \bot\alpha \rangle$ must be an initial segment of $\Gamma$, so that $\Gamma$ is a $\bot$-illegal run of $F^*$, contrary to our assumption. Thus, $\bot$ does not make any moves during the last iteration of LOOP either, and hence $\Gamma = \Theta_l$.

Now we can finish our proof of Lemma 7.2, i.e. show that $\mathbf{Wn}_e^{F^*}\langle\Gamma\rangle = \top$. If $\Gamma$ is a $\bot$-illegal run of $F^*$, we are done. Suppose now $\Gamma$ is not a $\bot$-illegal run of $F^*$. Then, by (3), $\mathbf{Wn}_e^{F^*}\langle\Gamma\rangle = \mathbf{Wn}_e^{L^*}\langle\rangle$. But, in view of (1), $\|L\|^*$ is a tautological combination of predicates and hence true at $e$, whence Lemma 6.3 implies that $\mathbf{Wn}_e^{L^*}\langle\rangle = \top$. Consequently, $\mathbf{Wn}_e^{F^*}\langle\Gamma\rangle = \top$. □

## 8 Completeness of CL1

In our completeness proof for **CL1** we employ the complementary logic **CL1**$'$ given by the following two rules:

(a) $\vec{H} \vdash F$, where $F$ is instable and $\vec{H}$ is the smallest set of formulas such that, whenever $F$ has a negative (resp. positive) surface occurrence of a subformula $G_1 \sqcap \ldots \sqcap G_n$ (resp. $G_1 \sqcup \ldots \sqcup G_n$), for each $i \in \{1,\ldots,n\}$, $\vec{H}$ contains the result of replacing this occurrence in $F$ by $G_i$.

(b) $H \vdash F$, where $H$ is the result of replacing in $F$ a positive (resp. negative) surface occurrence of a subformula $G_1 \sqcap \ldots \sqcap G_n$ (resp. $G_1 \sqcup \ldots \sqcup G_n$) by $G_i$ for some $i \in \{1,\ldots,n\}$.

**Lemma 8.1** **CL1** $\not\vdash F$ iff **CL1**$' \vdash F$ (any formula $F$).

**Proof.** We prove this lemma by induction on the complexity of $F$. It would be sufficient to verify the 'only if' part, for the 'if' part (which we do not need anyway) can be handled in a fully symmetric way. So, assume **CL1** $\not\vdash F$ and let us see that then **CL1**$' \vdash F$. There are two cases to consider:

*Case 1:* $F$ is stable. Then there must be a **CL1**-unprovable formula $H$ that is the result of replacing in $F$ some positive (resp. negative) surface occurrence of a subformula $G_1 \sqcap \ldots \sqcap G_n$ (resp. $G_1 \sqcup \ldots \sqcup G_n$) by $G_i$ for some $i \in \{1,\ldots,n\}$, for otherwise $F$ would be **CL1**-derivable by Rule **(a)**. By the induction hypothesis **CL1**$' \vdash H$, whence, by Rule **(b)**, **CL1**$' \vdash F$.

*Case 2:* $F$ is instable. Let $\vec{H}$ be the smallest set of formulas such that, whenever $F$ has a negative (resp. positive) surface occurrence of a subformula $G_1 \sqcap \ldots \sqcap G_n$ (resp. $G_1 \sqcup \ldots \sqcup G_n$), for each $i \in \{1,\ldots,n\}$, $\vec{H}$ contains the result of replacing this occurrence in $F$ by $G_i$. None of the elements of $\vec{H}$ is **CL1**-provable, for otherwise $F$ would be derivable in **CL1** by Rule **(b)**. Therefore, by the induction hypothesis, each element of $\vec{H}$ is **CL1**$'$-provable, whence, by Rule **(a)**, **CL1**$' \vdash F$. □

**Lemma 8.2** *If* **CL1** $\not\vdash F$, *then* $F$ *is not valid (any formula $F$).*

*In particular, if* **CL1** $\not\vdash F$, *then $F^*$ is not computable for some interpretation $^*$ that interprets all atoms as finitary predicates.*

**Proof idea.** Our proof of this lemma rests on a technique which, for a **CL1**-unprovable and hence **CL1**$'$-provable formula $F$, constructs an EPM $\mathcal{E}'$ and an interpretation $^*$ such that every HPM $\mathcal{H}$ loses the game $F^*$ against $\mathcal{E}'$ when $\mathcal{H}$ receives itself as an input. Precisely, "input $\mathcal{H}$" in this context means $\langle c, 0, 0, 0, \ldots\rangle$, where $c$ is the code of $\mathcal{H}$.

Revisiting our soundness proof for **CL1**, the central idea in it was to design $\top$'s strategy — the EPM $\mathcal{E}$ — following which guaranteed that the game would be eventually "brought down to" and stop (in the sense that no further moves would be made) at $L^*$ for some stable formula $L$. Such a strategy was directly extracted from a **CL1**-proof of $F$. In a perfectly symmetric way, a **CL1**$'$-proof of $F$ allows us to extract $\bot$'s strategy — the EPM $\mathcal{E}'$ — that makes sure that, in every legal scenario, $\top$'s play against $\mathcal{E}'$ over $F^*$ will be brought down to and stop at game $L^*$ for some instable formula $L$ — let us call such a formula $L$ the *limit formula* of the play. In view of Lemma 6.3, stopping at $L^*$ implies that $\top$ will be the loser as long as the predicate $\|L\|^*$ is false at the input. And the instability of $L$ means that, whatever the input $e$ is, $\|L\|^*$ is indeed false at $e$ for some $^*$, so that $\top$ loses $F^*$ on $e$ for such $^*$. However, the trouble is that $\top$'s different strategies — as well as different inputs $e$ of course — may yield different limit formulas $L$ and hence require different, perhaps conflicting, interpretations $^*$ to falsify $\|L\|^*$ at $e$. Since we are trying to show the non-validity rather than non-uniform-validity of $F$, the whole trick now is to find a *one* common interpretation $^*$ that, for $\top$'s arbitrary strategy (HPM) $\mathcal{H}$, would falsify $\|L\|^*$ at some input $e$, where $L$ is the limit formula that $\mathcal{H}$ and that very input $e$ yield. This is where a diagonalization-style idea comes to save our day. We manage to define an interpretation $*$ that makes the elementarization of every instable formula $G$ imply the following: "When HPM $\mathcal{H}$ plays against $\mathcal{E}'$ on input $\mathcal{H}$, the limit formula is not $G$". Now, if we let any given HPM $\mathcal{H}$ play on input $\mathcal{H}$ against $\mathcal{E}'$, the elementarization of the limit formula $L$ of the play — which, under interpretation $*$, claims that $L$ is not really the limit formula — is guaranteed to be false at $\mathcal{H}$. This ultimately translates into $\mathcal{H}$'s having lost the game $F^*$ on input $\mathcal{H}$.

**Proof.** Assume $\mathbf{CL1} \nvdash F$. Then, by Lemma 8.1, $\mathbf{CL1}' \vdash F$. Let us fix a particular $\mathbf{CL1}'$-proof of $F$, call it "the proof", and call formulas appearing in it "proof formulas". As in Section 7, we assume that every proof formula appears only once in the proof and comes with a fixed justification. We construct an EPM $\mathcal{E}'$ whose work is described as follows. At the beginning, this machine creates a record $E$ to hold proof formulas, initializes it to $F$, and then follows the following interactive algorithm:

**Procedure** LOOP: As long as $E$ is a proof formula, act depending on which of the two rules was used in the proof to derive $E$ from its premises:

**Case of Rule (a):** Keep granting permission until the adversary makes a move $\alpha$, then act depending on which of the following two subcases holds:

*Subcase (i):* $\alpha = \beta i$, where $\beta$ $E$-specifies a negative (resp. positive) surface occurrence of a subformula $G_1 \sqcap \ldots \sqcap G_n$ (resp. $G_1 \sqcup \ldots \sqcup G_n$) and $i \in \{1, \ldots, n\}$. Let $H$ be the result of substituting the above occurrence by $G_i$ in $E$. Then update (the content of) $E$ to $H$, and repeat LOOP.

*Subcase (ii):* $\alpha$ does not satisfy the condition of Subcase (i). Then go to an infinite loop in a permission state.

**Case of Rule (b):** Let $H$ be the premise of $E$ in the proof. $H$ is the result of substituting, in $E$, a certain positive (resp. negative) surface occurrence of a subformula $G_1 \sqcap \ldots \sqcap G_n$ (resp. $G_1 \sqcup \ldots \sqcup G_n$) by $G_i$ for some $i \in \{1, \ldots, n\}$. Let $\alpha$ be the $E$-specification of that occurrence. Then make the move $\alpha i$, update (the content of) $E$ to $H$, and repeat LOOP.

Note that the above description of $\mathcal{E}'$ is literally the same as the description of machine $\mathcal{E}$ given in Section 7, only with the words "positive" and "negative" interchanged. In view of the perfect symmetry between $\mathcal{E}$ and $\mathcal{E}'$, $\mathbf{CL1}$ and $\mathbf{CL1}'$, and clauses (a) and (b) of Lemma 6.4, arguing as in the previous section, we can conclude that, in each computation branch of $\mathcal{E}'$, LOOP will be iterated only a finite number of times and that $\mathcal{E}'$ is fair.

Let us fix some standard way of describing HPMs, and let

$$\mathcal{H}_0, \mathcal{H}_1, \mathcal{H}_2, \mathcal{H}_3, \ldots$$

be the list of all HPMs arranged according to the lexicographic order of their descriptions, so that each $c$ can be considered the code (or *a* code if we identify equivalent EPMs) of $\mathcal{H}_c$. Next, throughout the rest of this section, let $e_0, e_1, e_2, \ldots$ be the inputs defined by

$$e_0 = \langle 0, 0, 0, 0, \ldots \rangle, \ e_1 = \langle 1, 0, 0, 0, \ldots \rangle, \ \ldots, \ e_c = \langle c, 0, 0, 0, \ldots \rangle, \ \ldots.$$

For any $c \in \{0, 1, 2, \ldots\}$, let $B_c$ be the $(\mathcal{E}', e_c, \mathcal{H}_c)$-branch, $\Gamma_c$ the run cospelled by $B_c$, i.e. the $\mathcal{H}_c$ vs. $\mathcal{E}'$ run on $e_c$, and $L_c$ the value of the record $E$ of $\mathcal{E}'$ at the beginning of (and hence throughout) the last iteration of LOOP in $B_c$.

For an arbitrary $c \in \{0, 1, 2, \ldots\}$ and an arbitrary interpretation $*$, the following two statements can be proven in a way fully symmetric to the way we proved (1) and (3) in Section 7:

$$L_c \text{ is an instable proof formula.} \tag{4}$$

$$\text{Either } \Gamma_c \text{ is a } \top\text{-illegal run of } F^*, \text{ or } \mathbf{Wn}_{e_c}^{F^*}\langle \Gamma_c \rangle = \mathbf{Wn}_{e_c}^{L_c^*}\langle \rangle. \tag{5}$$

Let $G_1, \ldots, G_n$ be all the instable proof formulas. For each such $G_i$, let us fix a classical model $M_i$ (a true/false valuation for non-logical atoms extended to all $\sqcap, \sqcup$-free formulas in the standard classical way) in which $\|G_i\|$ is false. For each $i \in \{1, \ldots, n\}$, let $T_i$ be the predicate defined by

$$T_i \text{ is true at an input } \langle c, \ldots \rangle \text{ iff } L_c = G_i.$$

Now we define the interpretation $^*$ by stipulating that, for each atom $p$,

$$p^* = \vee\{T_i \mid 1 \leq i \leq n, \ p \text{ is true in } M_i\}.$$

($\vee S$ means the $\vee$-disjunction of all elements of $S$, understood as $\bot$ when $S$ is empty.)

Consider an arbitrary $c \in \{0, 1, 2, \ldots\}$. By (4), we must have $L_c = G_i$ for one of the $i \in \{1, \ldots, n\}$. Fix this $i$. It is easy to see that, for every atom $p$, $p$ is true in $M_i$ iff the predicate $p^*$ is true at $e_c$. This obviously extends from atoms to their $\neg, \wedge, \vee, \rightarrow$-combinations, so that the formula $\|G_i\|$ is true in $M_i$ iff the predicate $\|G_i\|^*$ is true at $e_c$. But, by our choice of $M_i$, the formula $\|G_i\|$, i.e. $\|L_c\|$, is false in $M_i$. Consequently, $\|L_c\|^*$ is false at $e_c$.[7] Then, by Lemma 6.3, $\mathbf{Wn}_{e_c}^{L_c^*}\langle\rangle = \bot$. Therefore, by (5), either $\Gamma_c$ is a $\top$-illegal run of $F^*$, or $\mathbf{Wn}_{e_c}^{F^*}\langle\Gamma_c\rangle = \bot$. In either case $\mathbf{Wn}_{e_c}^{F^*}\langle\Gamma_c\rangle = \bot$. But, in view of Lemma 6.2, $\Gamma_c$ is the run spelled by an $e_c$-computation branch of $\mathcal{H}_c$ — in particular, by the $(\mathcal{H}_c, e_c, \mathcal{E}')$-branch. Hence $\mathcal{H}_c$ does not win $F^*$: it loses this game against $\mathcal{E}'$ on input $e_c$.

Thus, no $\mathcal{H}_c$ computes $F^*$. This means nothing but that $F^*$ is not computable, because every HPM is $\mathcal{H}_c$ for some $c$. To officially complete our proof of the lemma, it remains to note that $p^*$ (any atom $p$) is finitary as it only depends on the first term $c$ of an input. $\square$

**Remark 8.3** The counterinterpretation $^*$ that we constructed in the proof of Lemma 8.2 for a **CL1**-unprovable formula $F$ interprets each atom $p$ as a predicate that depends only on the first term of an input and hence can be thought of as a unary *arithmetical predicate*, i.e a property (set) of numbers rather than a property of inputs. Let us for now agree to use the word "predicate" in this traditional sense. It is natural to ask the question about the arithmetical complexity of $p^*$. The answer we get is:

$$p^* \text{ is of complexity } \Delta_2 \quad (\text{any atom } p). \tag{6}$$

. Remember that an arithmetical predicate $P(c)$ is said to have complexity $\Sigma_2$ iff it can be written as $\exists x \forall y Q(c, x, y)$ for some decidable predicate $Q(c, x, y)$; and $P(c)$ is of complexity $\Delta_2$ iff both $P(c)$ and $\neg P(c)$ are of complexity $\Sigma_2$.

Our proof of Lemma 8.2 defines $p^*$ as the disjunction of some $T_i$, that we now think of as unary arithmetical predicates and write as $T_i(c)$. Disjunction is known to preserve $\Delta_2$ — as well as $\Sigma_2$ — complexity, so, in order to verify (6), it would be sufficient to show that each $T_i(c)$ ($1 \leq i \leq n$) is of complexity $\Delta_2$. Looking at the meaning of $T_i(c)$, this predicate asserts nothing but that the value of record $E$ in branch $B_c$ will stabilize at $G_i$. So, $T_i(c)$ can be written as $\exists x \forall y (y \geq x \rightarrow K_i(c, y))$, where $K_i(c, y)$ means "the value of record $E$ at the $y$th computation step of branch $B_c$ is $G_i$". Furthermore, in view of (4), the value of $E$ should indeed stabilize at one of the instable proof formulas $G_1, \ldots, G_n$. Hence, $\neg T_i(c)$ is equivalent to $\vee\{T_j(c) \mid 1 \leq j \leq n, j \neq i\}$. Consequently, in order to show that each $T_i(c)$ is of complexity $\Delta_2$, it would suffice to show that each $T_i(c)$ is of complexity $\Sigma_2$. For the latter, in turn, verifying that $K_i(c, y)$ is a decidable predicate would be sufficient. But $K_i(c, y)$ is indeed decidable. A decision procedure for it first constructs the machine $\mathcal{H}_c$ from number $c$. Then it lets this machine play against $\mathcal{E}'$ on input $\langle c, 0, 0, 0, \ldots \rangle$ as described in the proof idea for Lemma 6.2. In particular, it traces, in parallel, how the configurations of the two machines evolve up to the $y$th computation step of $\mathcal{E}'$, i.e. its $y$th configuration. Then the procedure looks at the value of record $E$ in that configuration, and says "yes" or "no" depending on whether the latter is $G_i$ or not.

---

[7] We have just shown that, as long as $G_i = L_c$, $\|G_i\|^*$ is false at $e_c$. In other words, (truth of) $\|G_i\|^*$ (at $e_c$) implies $L_c \neq G_i$. This is exactly what was meant in 'Proof idea' by saying that $^*$ makes the elementarization of every instable formula $G$ imply "When $\mathcal{H}$ [i.e. $\mathcal{H}_c$] plays against $\mathcal{E}'$ on input $\mathcal{H}$ [i.e. $e_c$], the limit formula [i.e. $L_c$] is not $G$".

# Index